\documentclass[10pt,a4paper,twoside]{aphs}
\usepackage{amsfonts,amssymb}
\usepackage{graphicx}

\def\LaTeX{L\kern -.36em\raise .3ex\hbox{\sc a}\kern -.15em T\kern -.1667em%
\lower .7ex\hbox{E}\kern -.125em X}

 4

\font\erm=cmr8
\font\eit=cmti8

\begin{document}

\thispagestyle{empty}

\begin{flushright}
{\eit Acta Physicae Superficierum} {\erm $\bullet$ Vol VI $\bullet$ 2005}
\end{flushright}



\title{ELECTRON TRANSPORT THROUGH \\ THE TWO-DIMENSIONAL ATOMIC LATTICE
}

\author{T. Kwapi\'nski, Z. Korczak and M. Ja\L ochowski}
\address{Institute of Physics, M. Curie-Sk\l odowska University \\
Pl. M. Curie-Sk\l odowskiej 1, 20-031 Lublin, Poland}

\maketitle

\vspace{0.2cm}

\abstract{The electrical conductivity changes of the Si(111)-(6x6)Au surface at early stage of Pb deposition was studied
experimentally and theoretically as a function of coverage. Pb deposition onto a Si(111)-(6x6)Au surface induce strong change of
the conductance even at a low coverage ($<1$ monolayer). The experimental results are analyzed using the theoretical model of
two dimensional (2D) square atomic lattice.}

\section{\label{sec1} Introduction }

The electrical conduction through semiconducting surface layer is very sensitive to the presence of individual foreign atoms.
 These atoms act predominantly as dopants and can, depending on the semiconductor type, enhance or decrease the electrical
 conductivity of the depletion or accumulation layer \cite{has}. Presence on the semiconductor
 surface of a monoatomic metallic layer allows to separate these quasi two-dimensional semiconducting effects from truly two-dimensional
 phenomena arising during interaction of individual atoms with the surface. In this case the conventional theoretical description
 which takes into account the statistics of current carriers in semiconductors or the well known Drude conductivity theory, is not
 sufficient.

In this study we propose a theoretical model which allows to calculate the electrical conductance of truly two-dimensional
lattice of atoms with single atoms deposited on it. The model simulates the initial stage of well known phenomenon of
oscillating electrical resistivity observed during layer by layer growth of metallic layer \cite{jal}. The model is illustrated
by experimental data of the electrical conductance of the Si(111)-(6$\times$6)Au system, measured during deposition of
individual atoms of Pb on it. The morphology of the surface covered with Pb atoms is studied with scanning tunnelling microscope
(STM).

\section{\label{sec1}Experimental setup and results }

The measurements were performed in a UHV chamber with a base pressure of less than 1x10$^{-10}$ mbar. The structure of the
substrate and the deposition of Pb were monitored by reflection high energy electron diffraction (RHEED) system. An n-type
Si(111) wafer of around 25 $\Omega$cm resistivity and 18x4x0.4 mm$^3$ size was used as substrate. Electrical conductivity was
measured in situ by four-point probe method. An alternating current: I = 2 $\mu$A, 17 Hz was sent through the outer-most Ta
clamps contacts, while the ac voltage was measured across the inner two W wires kept in elastic contact with the wafer. Before
each measurement run, the surface was cleaned to obtain a clear Si(111)-7x7  RHEED pattern, by few flash heating. In order to
prepare the Si(111)-(6x6)Au surface structure, 1.3 ML of Au were deposited on Si(111)-7x7 superstructure. Annealing for 1 min at
about 950 K and slow cooling to room temperature (10K/min.) resulted in the appearance of a sharp (6x6)Au superstructure RHEED
pattern. During deposition of Pb the sample was kept at 78 K. The conductance could be measured simultaneously with deposition.
The amount of deposited material in units of monolayer (ML = 7.8x10$^{14}$ atoms/cm$^2$) was monitored with a quartz crystal
oscillator.

\begin{figure}[h]
\begin{center}
 \resizebox{0.7\columnwidth}{!}{
  \includegraphics{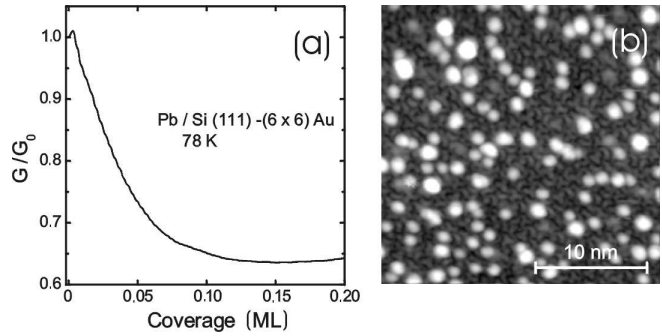}}
\end{center}
 \caption{\label{Fig1} (a) The relative conductance changes recorded during Pb deposition
  onto a Si(111)-(6x6)Au surface at 78 K. $G_{0}$ denotes the bare substrate conductance.
  (b) STM topographic image of Si(111)-(6 $\times$ 6) substrate covered with 0.05 ML of Pb at 102 K.
   The smaller round bright species are the Pb single atoms.}
\end{figure}
Fig.1a shows the relative conductivity changes during Pb deposition at 78 K onto a Si(111)-(6x6)Au surface for coverage up to
0.2 ML. The decrease of the conductivity starts from the beginning of the deposition. The conductivity reaches a minimum at
around 0.12 ML coverage and afterwards it increases. An abrupt increase of the conductivity at opening of the shutter is
connected with an influence of radiation from evaporator. The samples studied with the STM were prepared in similar condition in
another UHV system equipped with Omicron VT STM.  Pb films of various thickness have been deposited on such prepared substrates
 mounted in the cooled STM stage. During the deposition the substrate temperature was equal to about 100 K. The pressure of
 the chamber during the Pb deposition was below $7\times10^{11}$ mbar. Fig. 1b shows STM scan taken after deposition of about
 0.05 ML of Pb.  Previous STM studies revealed \cite{jal2}  that at this coverage the
 visible small features are predominantly single Pb atoms. A few larger and brighter species are probably composed of more than
 single atom. The density of these atom clusters increased with Pb coverage and finally, for coverage corresponding to 1 ML of Pb(111),
  the whole surface was covered with epitaxial monoatomic layer of Pb.


\section{\label{sec1}Theoretical description and results}

\begin{figure}[h]
\begin{center}
 \resizebox{0.75\columnwidth}{!}{
  \includegraphics{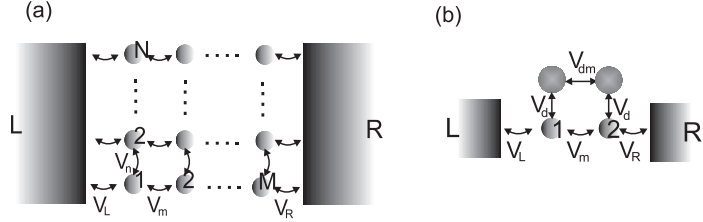}}
\end{center}
 \caption{\label{Fig2} (a) The top view of the system of two-dimensional square lattice of atoms ($N\times
 M$) connected to the left (L) and right (R) leads. The additional atoms are deposited on the top of this
 lattice;   (b) Side view of 2-atom chain with two adatoms.
$V_n, V_m, V_{L}, V_R$, $V_d$ and $V_{dm}$ represent the hybridization matrix elements  responsible for electron transport
between corresponding atoms.}
\end{figure}

In this section we study theoretically the conductance through a model of two-dimensional (2D) square atomic lattice. The model
is schematically shown in Fig. \ref{Fig2}a. The central region, between the left (L) and right (R) electron reservoirs, consists
of an array of $N\times M $ atom sites. Additional atoms (adatoms) are added on the top of 2D lattice and cause the disturbance
of the electron transport through the system. A group of coupled adatoms can be treat as a cluster of atoms. The 2D lattice of
atoms is considered as the effective substrate which corresponds to, used in the experiment, Si(111)-(6x6)Au substrate. The
additional atoms correspond to deposited Pb atoms and the left and right leads represent the voltage electrodes. We assume here,
that each adatom is coupled only with the single substrate atom.  Moreover, 3D island growth is not considered.

The total tight-binding Hamiltonian for the system can be
decomposed as $H=H_0+H_1+\widetilde{V}_1$ where $H_0$ describes
electron states in the left and right electrodes, in the substrate
and also interactions between substrate atoms. The adatom single
electron states are described by the Hamiltonian $H_1$.
$\widetilde{V}_1$ concerns adatom-adatom and substrate atom-adatom
interactions. For simplicity, a single orbital per atomic site is
considered. Only nearest-neighbor interactions are taken into
account. We consider the total Hamiltonian $H$ in standard
second-quantized notation in the following form:

\begin{eqnarray}
 H_0 &=& \sum_{\vec k\alpha=L,R} \varepsilon_{\vec k\alpha}
a^+_{\vec k\alpha} a_{\vec k\alpha} + \sum_{i=1}^N \sum_{j=1}^M
\varepsilon_{ij} a^+_{ij} a_{ij}+ \sum_{i=1}^N  \sum_{\vec kL,\vec
kR} V_{L/R} a^+_{\vec kL/\vec kR}a_{i1/M} \nonumber\\ &+&
\sum_{i=1}^{N} \sum_{j=1}^{M-1} V_m a^+_{ij} a_{ij+1}+
\sum_{i=1}^{N-1} \sum_{j=1}^{M} V_n a^+_{ij} a_{i+1j}+{\rm h.c.}
 \label{eq1}
\end{eqnarray}
\begin{eqnarray}
 \widetilde{V}_1 &=& \sum_{i=1}^N  \sum_{j=1}^M V_{d}
c^+_{ij}a_{ij}\Delta_{ij} + \sum_{i=1}^N \sum_{j=1}^{M-1} V_{dm}
c^+_{ij}c_{ij+1}\Delta_{ij}\Delta_{ij+1} \nonumber\\&+&
\sum_{i=1}^{N-1} \sum_{j=1}^{M} V_{dn}
c^+_{ij}c_{i+1j}\Delta_{ij}\Delta_{i+1j}+{\rm h.c.}
 \label{eq3}
\end{eqnarray}
and $H_1 = \sum_{i=1}^N \sum_{j=1}^M \varepsilon_{ij}^d c^+_{ij}
c_{ij} \Delta_{ij}$. Here the operators $a_{\vec
k\alpha}(a^+_{\vec k\alpha})$, are the annihilation (creation)
operators for the reservoirs conduction electron with the wave
vector $\vec k$, ($\alpha = L, R$). $a_{ij}(a^+_{ij})$ represents
the annihilation (creation) operators for the electrons localized
on the substrate atom $ij$ ($i=1,...,N$, $j=1,...,M$) and
$c_{ij}(c^+_{ij})$  on adatoms above $ij$ substrate atom.
$\varepsilon_{ij}$ and $\varepsilon_{ij}^d$ are the single
electron energy levels of the substrate atoms and adatoms,
respectively. The elements $V_{\vec kL(R)}$ are the hybridization
matrix elements responsible for electron transport between
corresponding substrate atoms and electrodes; $V_{n(m)}$ between
the lattice atoms; $V_{dn (dm)}$ between adatoms and $V_d$ -
between adatom and the 2D lattice atom. The function $\Delta_{ij}$
in the Hamiltonian is equal to $1$ if there is an additional atom
at site $ij$. Otherwise it is equal to zero. This function allows
to control number of adatoms included into calculations. Note,
that the first term in Eq. 2 describes the interaction between the
adatom  and the substrate atom at site $ij$.

Electron and transport properties are analyzed within the framework of the Green's function technique. In the energetic
representation Green's function satisfies the following equation of motion \cite{dat}:
\begin{eqnarray}
E G^r_{\beta_1 \beta_2}(E)= \langle[a_{\beta_1};a^+_{\beta_2}]_+\rangle+\langle [\{a_{\beta_1},H]; a^+_{\beta_2}\}\rangle_E
\label{eq5}
\end{eqnarray}
Here $E$ is the energy and $\beta_{i}$ represents the
corresponding atomic site. The knowledge of the appropriate
Green's function allows to obtain the electron local density of
states (LDOS) of the substrate atoms or adatoms, i.e. for the site
$\beta$ one can write: $LDOS_{\beta}=-{1\over \pi} {\rm Im}
G^r_{\beta,\beta}(E) $. The zero-temperature linear conductance,
obtained for zero or very small voltage, is given in terms of the
Green function by the following expression: $G(E) = {2e^2\over
h}T(E)$, where $T(E)={2e^2\over h} {\rm Tr \{\Gamma_L G^r \Gamma_R
(G^r)^*\}}$ is the transmittance of the system. The matrix
$\Gamma_L$ is expressed as follows:
$(\Gamma_L)_{ij,kl}=\Gamma^{L}\delta_{i1}\delta_{k1}$, where $
\Gamma^{L}=2\pi \sum_{\vec kL} |V_{\vec kL}|^2
\delta(\varepsilon-\varepsilon_{\vec kL}) =2\pi {V^2_{L}/ D}$.
Here $D$ is the effective bandwidth of the left lead density of
states. The similar expression we can write for $\Gamma_R$ matrix,
i.e. $(\Gamma_R)_{ij,kl}=\Gamma^{R}\delta_{iM}\delta_{kM}$. Using
above relations the conductance of the system becomes:
\begin{eqnarray}
G(E) &=& {2e^2\over h}\Gamma^L\Gamma^R \left| \sum_{i=1}^N
\sum_{j=1}^M G^r_{1i,Mj}(E)\right|^2
 \label{eq10}
\end{eqnarray}
Note, that the conductance is expressed only by the Green's functions $G^r_{1i,Mj}(E)$ between atoms in the first and the last
column, cf. Fig. \ref{Fig2}a (adatoms are not connected to the leads). The appropriate Green functions are obtained using Eq. 3
and the Hamiltonian. The general matrix equation for $G^r_{ij,kl}(E)$ can be written in the form:  $A \cdot G^r=I$, where $I$ is
the unit matrix. For the substrate atoms the matrix $A$ is expressed as follows:
\begin{eqnarray}
A_{ij,kl}(E)&=&(E-\varepsilon_{ij})\delta_{ik}\delta_{jl}+i{\Gamma^L
\over 2} \delta_{1j}\delta_{1l}+i{\Gamma^R
\over 2} \delta_{Mj}\delta_{Ml} \\
&&- V_m(\delta_{ik}\delta_{j+1 l}+\delta_{ik}\delta_{j-1 l})
-V_n(\delta_{i+1 k}\delta_{jl}+\delta_{i-1 k}\delta_{jl})-V_d
\Delta_{ij} \nonumber
 \label{eq11}
\end{eqnarray}
and similar for the adatoms.
The retarded Green functions are obtained by
finding the inverse of the matrix $A$, i.e. $
G^r_{ij,kl}(E)=A^{-1}_{ij,kl}(E)$.
%


Next, we apply the above formalism to the case of two-atom lattice connected to the left and right leads and two adatoms as
shown Fig. \ref{Fig2}b. This simple model allows to study the influence of the couplings between adatoms and also between these
atoms and the substrate atoms on the LDOS and conductance through the system.
We treat the Fermi energy of the system as the reference energy point i.e. $E_F=0$ (the linear conductance is obtained for the
left and right chemical potentials  equal to the Fermi level, $\mu_L=\mu_R=E_F$), and take into consideration the case of
$\Gamma^L=\Gamma^R=\Gamma$. Moreover, the same electron energy of all substrate atoms and adatoms are assumed i.e.
$\varepsilon_{ij}=\varepsilon_0$, $\varepsilon_{ij}^d=\varepsilon_d$. This assumption is quite reasonable as only linear regime
is considered. All energies are given in $\Gamma$ units and the conductance in $2e^2/h$ units.

The transport properties of the system are dependent on the density of states of this system. The conductance is proportional to
the DOS of the substrate at the Fermi level \cite{dat}. For the considered system the matrix $A$ can be written in the form:
\begin{eqnarray}
A=\left( \begin{array}{cccc}
  E-\varepsilon_0+i{\Gamma/2} & -V_m                        & -V_{d}          & 0     \\
  -V_m                        & E-\varepsilon_0+i{\Gamma/2} & 0               & -V_{d} \\
  -V_d                        & 0                           & E-\varepsilon_d & -V_{dm} \\
  0                           & -V_d                        & -V_{dm}         & E-\varepsilon_d \\
\end{array}
\right)  \label{eq13}
\end{eqnarray}
By calculating the inverse of the matrix $A$ we find the
appropriate Green's function needed to obtaining the LDOS. For the
sake of symmetry of the system the LDOS of the substrate atoms and
adatoms are equal, i.e. $LDOS_1=LDOS_2$ and $LDOS_3=LDOS_4$ - the
indexes 1 and 2 (3 and 4) correspond to the substrate atoms
(adatoms).
\begin{figure}[h]
\begin{center}
 \resizebox{0.5\columnwidth}{!}{
  \includegraphics{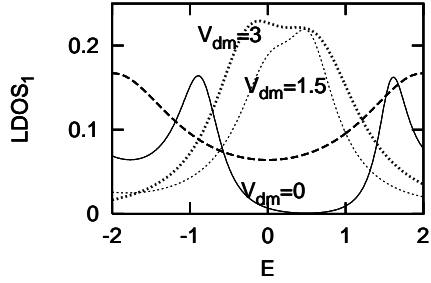}}
 \end{center}
 \caption{\label{Fig3} The LDOS of the substrate atoms for the system
 schematically shown in Fig. \ref{Fig2}b. The couplings between atoms are: $V_d=2$, $V_m=2$
 and $V_{dm}=0, 1.5, 3$ - solid, dotted and doted-thick lines, respectively.
 The dashed line corresponds to the case:
 $V_m=2$, $V_d=0$ (no interaction between adatoms and the substrate atoms), $\Gamma=1$.
 }
\end{figure}
The influence of the coupling between adatoms, $V_{dm}$, on the LDOS of the substrate atoms is presented in Fig. \ref{Fig3}. The
dashed line corresponds to the case $V_d=0$ - no interaction between adatoms and the substrate atoms. The others curves
correspond to the coupling between adatoms:
 $V_{dm}=0, 1.5, 3$ - solid, dotted, dotted-thick lines
and $V_d=V_m=2$. The electron energy levels are $\varepsilon_0=0$ and $\varepsilon_d=0.5$ (we assume that the adatom level, in
comparison to the substrate-atom level, possesses the higher energy because the ionization energy of Au is greater then for Pb
atom). The dashed curve possesses two maxima for $E\simeq \pm V_{m}$ which is known behaviour of atomic wires e.g. \cite{yam}.
The additional atoms blockade the LDOS around the adatom energy level $E=\varepsilon_d$,  cf. the solid line. For increasing
value of the coupling $V_d$ the LDOS dip broadens (not shown here). However, for nonzero coupling strength between adatoms the
LDOS increases in the regime of energies around the Fermi energy (so about $\varepsilon_d$ and $\varepsilon_0$) - cf. the dotted
lines in Fig. \ref{Fig3}. It causes that the conductance of the system increases with increasing $V_{dm}$.

The main conclusions from this simple model are: (i) non-coupled adatoms on the 2D lattice cause decreasing of the LDOS of the
substrate (and also the conductance) obtained for the Fermi energy while (ii) the adatoms which are coupled with each other
cause increasing of the LDOS at the Fermi level (so the conductance through the system is greater then for non-coupled atoms in
the cluster).

Now we take into consideration the system of the two-dimensional
substrate of $N\times M$ atoms and adatoms deposited on it. A
series of calculations were performed for the system with
increasing number of adatoms. These adatoms are placed on the
lattice randomly and the conductance is obtained numerically.
\begin{figure}[h]
\begin{center}
 \resizebox{0.85\columnwidth}{!}{
  \includegraphics{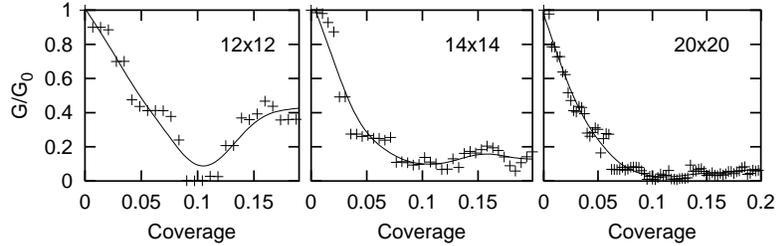}}
\end{center}
 \caption{\label{Fig4} The relative conductance for the lattice of atoms:
$N=M=12$ (left panel), 14 (middle panel), 20 (right panel) versus the coverage of the substrate. The other parameters are:
$V_d=2$, $V_n=V_{m}=3$, $V_{dn}=V_{dm}=5$ and $\varepsilon_0=0$, $\varepsilon_d=0.5$. }
\end{figure}
Fig.~\ref{Fig4} shows the relative conductance $G/G_0$ ($G_0$ is the conductance of the substrate without adatoms) for the
lattice of atoms: $N=M=12$ (left panel), 14 (middle panel), 20 (right panel) versus the coverage of the substrate. The lines are
plotted for better visualization and only the coverage of order $20\%$ is considered as the time computation increases very
strongly with increasing number of atoms in the system. For greater number of atoms in the 2D lattice the greater number of
points is observed (each cross corresponds to one adatom deposited on the substrate). At the beginning the conductance decreases
with increasing coverage of the substrate. This is due to the single atoms deposited on the lattice (cf. also Fig. \ref{Fig3}).
However, for the substrate coverage about 0.12 ML, the conductance possesses the local minimum. In that point the first cluster
of adatoms appears and for greater coverage more clusters are observed. These clusters are responsible for increasing the
conductance although there is no percolation between the left and right electrodes.
This effect is also visible in Fig. \ref{Fig3} for the one-dimensional case. Note, that the results presented in Fig. \ref{Fig4}
could be somewhat different for various distribution of the addatoms on the substrate. But it was checked that the general
results remain unchanged.

To conclude, the theoretical results show that the conductance of 2D atomic lattice decreases with increasing coverage. For
greater coverage ($>0.1$ML), due to clusters of adatoms, the local minimum in the conductance appears. These results are in good
agreement with the experiment described in the second section and also with the STM measurements.  Due to the simple theoretical
model the results presented in this section should be considered only as the qualitative conclusions.

\subsection*{Acknowledgements}

The work is partially supported by the 2004 UMCS Grant and the Foundation for Polish Science.


\end{document}